\documentstyle[12pt]{article}
 
\begin{document}
 
\title{PARTON DISTRIBUTION WITHIN VIRTUAL PARTICLES
AND\enspace  $\gamma^{*}p$ \enspace INTERACTION}
 
\author{ I. Royzen }
 
\centerline{Lebedev Physical Institute, 53 Leninski Prospect, Moscow 117333,
Russia}
 
\begin{abstract}
A simple relativistic model is suggested that elucidates the
qualitative difference in valence quark distributions of bosons in ground and
virtual states. The inelastic diffraction of highly virtual photon in DIS is
discussed.
\end{abstract}
 
The unambiguous experimental evidence has been obtained ~\cite{aa} in DIS
that the cross section of inelastic diffraction \enspace $\gamma
^{*}(Q^2)+p\rightarrow hadrons+p$ \enspace is unexpectedly large even at
$Q^2\gg 1\mbox{ $GeV$}^2$ and that its fraction in the total cross section of
\enspace $\gamma ^{*}p$\enspace interaction is almost independent of $Q^2$ .
These results indicate the essential role which play fluctuations of highly
virtual space-like photon into the hadronic states with sufficiently large
transverse size. I would like to present a very simple and rather toy model
which demonstrates that the transverse size of any highly virtual boson
treated as a function of two variables, $Q^2$ and $x$ ($x$ being the energy
fraction of a parton), is really expected to peak sharply near $x=m_q^2/Q^2$,
approaching that of the boson on the mass shell. Therefore, this parton
configuration should produce the diffraction pattern in the scattering of
highly virtual boson similar to that of the real one. Being the striking
qualitative effect, this feature of virtual particle interaction is expected
to manifest itself (though, maybe, being attenuated numerically) within the
frameworks of more realistic approaches. The consideration is based on the
dispersion relation ideology ~\cite{bb} that parton distribution within the
real or space-like boson (to be certain, I refer below to the photon) should
be expressed by an integral over the time-like $q\bar q$-fluctuations in the
relevant channel.\footnote {A similar
ideology has been exploited in the paper ~\cite{cc} for the evaluation of quark
distribution at rather low values of $Q^2$.} These fluctuations incorporate
the low laying resonances ($\rho _0,\omega ,etc.$) as well as the continuum
background which contributes predominantly to the parton distribution at
large $Q^2$. It is easy to estimate the weight function attached to a
certain $q\bar q$-fluctuation. The contribution of $q\bar q$-fluctuation
with the mass $M\geq 2m_q$, $m_q$ being the quark mass, is proportional to
the time $\Delta {t}$ which this fluctuation spends in the state with
four-momentum squared equal to $(-Q^2)$. In turn, for the photon with
momentum $\vec P^2\gg Q^2$,\enspace
$\vec P^2\gg M^2$
\begin{eqnarray}\label{1} \Delta{t}\enspace \approx\enspace
\frac{1}{\sqrt{{\vec P}^2 + M^2}-\sqrt{{\vec P}^2 - Q^2}}\enspace
\simeq\enspace \frac{2\vert{\vec P}\vert}{M^2 + Q^2} \end{eqnarray} in the
accordance with the energy-time uncertainty relation. It follows from Eq.(1)
that parton distribution within the real photon $Q^2=0$ is saturated
predominantly by resonances of the lowest masses only ( namely, by $\rho
_0,\omega $ and $\phi $ because the mass squared of the next resonance
situated at the 1st daughter Regge trajectory is about four times larger),
while for the photon with $Q^2\gg\nolinebreak 1\mbox{ $GeV$}^2$ it is
contributed essentially by the whole set of $q\bar q$-fluctuations with
$M^2\leq Q^2$.  Thus, the calculation of quark distribution within virtual
photon is reduced to summing up the properly weighed quark distributions in\\
$q\bar q$ -fluctuations with $2m_q^2\leq M^2$. For the moment, the many
particle (evolution) aspect of the problem is put aside. It will be briefly
discussed below.
 
The simplest model is considered to describe $q\bar q$-fluctuations: two
spinless particles of the same mass $m_q$ (below referred as ''quarks'')
interact via the confining potential
\begin{eqnarray}\label{2}
U = 0, \hspace{1cm} (x_1 - x_2)^2 + (y_1 - y_2)^2 + \frac{(z_1 - z_2)^2}
{1 - v^2}\enspace <\enspace R^2,
\end{eqnarray}
$$
U=+\infty ,\hspace{4cm}outside\enspace  of\enspace  this\enspace  ellipsoid,
$$
where $x_1,y_1,z_1$ and $x_2,y_2,z_2$ are the quark coordinates, $R$ is a
constant, $1\mbox{ $GeV$}^{-1}<R<5\mbox{ $GeV$}^{-1}$, and $v$ is the
velocity of their CMS along the axes $z$. The corresponding Shrodinger
equation reads
\begin{eqnarray}\label{3} (\enspace\sqrt{{\vec p}^2_1 + m^2_q} +
\sqrt{{\vec p}^2_2 + m^2_q} - \frac{M}{\sqrt{1 - v^2}}\enspace)\enspace
{\psi(\vec p_1,\vec p_2)}\enspace=\enspace 0 \end{eqnarray} The solution of
Eq.(3) is obviously proportional to $\delta $-function of the expression in
brackets.  In the limit $v\rightarrow 1,\enspace (|{\vec P}|\simeq |{P_z}|\to
\infty )$ \begin{eqnarray}\label{4} \sqrt{{\vec p}^2_1 + m^2_q} +\sqrt{{\vec
p}^2_2 + m^2_q}\enspace \simeq\enspace\frac{m_T} {\sqrt{x(1 - x)(1 - v^2)}}
\end{eqnarray} where $x=|{\vec p_1}|$ / $(|{\vec p_1}|+|{\vec p_2}|)$ and $m_T$
is the, so-called, transverse mass, \begin{eqnarray}\nonumber m_T\enspace
=\enspace \sqrt{p^2_{1x} + p^2_{1y} + m^2_q}\enspace =\enspace \sqrt{p^2_{2x} +
p^2_{2y} +m^2_q}\enspace \equiv\enspace \sqrt{p^2_T + m^2_q} \end{eqnarray}
Thus, at $Q^2\gg 1 \mbox{ $GeV$}^2$ the wave function of highly
virtual photon $\gamma ^{*}(Q^2)$ is approximately equal to the integral over
$M$ of these $\delta $-functions weighed in the accordance with Eq.(1), what
leads to \footnote{Strictly speaking, one has to sum up the projections of the
solutions of Eq.(3) to the state with $J = 1$, however it does not affect the
result, since \\$\cos{\theta}$ = $(p_{1z} + p_{2z})$ / $(\vert{p_1 + p_2}\vert)
\to 1$} \begin{eqnarray}\label{5} \psi_{\gamma^*}\enspace\simeq\enspace
\frac{A \vert{\vec P}\vert\sqrt{1 -v^2}}{Q^2 + m^2_T/x(1 - x)}, \end{eqnarray}
A being the normalization constant. Since the constraints $p_{1z}+p_{2z}=P_z$
and $\vec p_{1T}+\vec p_{2T}=0$ are to be allowed for, $$ |{\psi _{\gamma
^{*}}}|^2\enspace \equiv \enspace \frac{dW}{d^3{\vec p}_1d^3{\vec
p}_2}\enspace\sim \enspace \frac{dW}{dm_T^2d(p_{1z}-p_{2z})} $$ The standard
relativistic kinematics ~\cite{dd} gives at $m_T^2\ll {\vec p}_1^2,\enspace
m_T^2\ll {\vec p}_2^2$ $$ d(p_{1z}-p_{2z})\enspace
=\enspace\frac{m_Tdx}{2x^{3/2}(1-x)^{3/2}\sqrt{1-v^2}}$$ and, finally, the
normalized quark distribution reads as \begin{eqnarray}\label{6}
\frac{dW}{dm_Tdx}\enspace \simeq\enspace {\frac {4Q}{\pi}} \frac{m^2_T}
{x^{3/2}(1 - x)^{3/2} \lbrack Q^2 + m_T^2/x(1 - x)\rbrack^2} \end{eqnarray}The
most interesting feature of this distribution manifests itself, if one
estimates the mean transverse size of the photon as the function of $x$ and
$Q^2$ \begin{eqnarray}\label{7} <r(x,Q^2)> \simeq <\frac{1}{m_T}> = \int
\limits_{m_q}^{\infty} {\frac{dm_T}{m_T} \frac{dW}{dm_Tdx}}\simeq
\frac{2x^{-1/2} (1 - x )^{-1/2}}{\pi Q\lbrack {1 + m^2_q/Q^2x(1-x)}\rbrack}
\end{eqnarray}  Herefrom, the value of  $<r(x,Q^2)>$ is peaked sharply near the
points \begin{eqnarray}\label{8} x = \frac{m^2_q}{Q^2}\qquad and\qquad (1 - x)
= \frac{m^2_q}{Q^2} \end{eqnarray} {\em Irrespectively of $Q^2$} it is equal
there to \begin{eqnarray}\label{9}  r_{max}\enspace \simeq\enspace (\pi
m_q)^{-1}\enspace \geq\enspace 1 \mbox{ $GeV$}^{-1} \end{eqnarray} The $q\bar
q$- -fluctuations of such a large size should exhibit the hadron-like strong
interaction with the cross section which is caused by the degree of color
descreening, i.e., is proportional to $<r^2(x,Q^2)>$. If the color descreening
gets working effectively within the range $r_0\leq r\leq r_{max}$
(in the accordance with Eq.(7), at\enspace $(r_{max}/r_0)\geq 2$\enspace it is
approximately equivalent to \\$(\pi r_0m_q^2)^2/4Q^2=x_1\leq x(1-x)\leq
x_2=4/(\pi r_0Q)^2$), then this cross section\begin{eqnarray}\label{10}
\tilde\sigma_{\gamma^*p}\enspace \approx\enspace 4\pi\alpha\enspace
\frac{8\pi}{\pi^2 Q^2} \int\limits_{x_1}^{x_2} \frac{dx}{x}\enspace
\approx\enspace 4\pi\alpha\enspace \frac{8}{\pi{Q^2}}[2{\ln
\frac{2}{\pi{r_0m_q}}-1]}\enspace \geq\enspace \frac{4\pi\alpha}{Q^2}
\end{eqnarray} Thus, the contribution of the large size (soft) $\gamma ^{*}p$
interaction turns out to be of the order (or even larger), than that of the
''normal'' DIS point-like $\gamma ^{*}p$ interaction ($\simeq 4\pi \alpha
/Q^2$). As to the former one, it is reasonable to expect that it should
resemble closely the features of interaction of the real photon which exhibits
a diffraction pattern with a certain non-vanishing (as the energy increases)
fraction of photon inelastic diffraction into hadrons.  The above model is far
from being realistic and hardly can be used for obtaining the reliable
quantitative results. However, its striking qualitative features can, most
probably, stand against the necessary corrections which are of two kinds.
First, in the framework of the two-particle approximation one has to allow for
the direct quark-antiquark interaction via gluonic exchange and, therefore, to
consider the linear confining potential instead of the potential wall of the
infinite height. The corresponding corrections diminish the asymmetry in $q$
and $\bar q$ energies and, hence, the contribution of the large size quark
configurations in the $\gamma ^{*}$ wave function. Second, the many particle
aspect of the problem is to be taken into account which is associated with
the evolution equation or gluon bremsstrahlung (especially important is
that produced by the slow quark or antiquark).  These processes enlarge
crucially the population of low energy partons and should lead to the strong
enhancement of the large size effects.  Thus, it seems reasonable to consider
the distributions (6) and (7) as a quite natural initial conditions for the
calculation of realistic parton distribution within highly virtual space-like
photon.\\
 
The work is supported, in part, by Russian Foundation for Basic Researches,
grant No: 96-02-19572.\\

\end{document}